\def\i{{\rm i}}
\def\d{{\rm d}}
\documentstyle[12pt]{article}

\begin{document}
\pagestyle{empty}


\vspace*{10\baselineskip}

\noindent
{\bf NONCOMMUTATIVE DYNAMICS}\\
{\it Dedicated to L.C. Biedenharn}\vspace{3\baselineskip}

\noindent
\hspace*{1in}
Jakub Rembieli\'nski\vspace{\baselineskip}

\noindent
\hspace*{1in}
University of {\L}\'od\'z\\
\hspace*{1in}
Department of Theoretical Physics\\
\hspace*{1in}
ul.\ Pomorska 149/153\\
\hspace*{1in}
90--236 {\L}\'od\'z, Poland\vspace{3\baselineskip}

\noindent
{\bf INTRODUCTION}\vspace{\baselineskip}

The first step in the noncommutative dynamics was undertaken by
L.C. Biedenharn$^1$ who considered the quantum noncommutative
harmonic oscillator. Recently Aref'eva and Volovich$^2$
published paper devoted to some nonrelativistic dynamical system
in a noncommutative phase-space framework.

Noncommutative analogon of the Galilean particle, as described
in Aref'eva and Volovich$^2$, has two main features:

-- Consistency of the formalism demands noncommutativity of the
inertial mass. This phenomena holds also in Rembielinski$^3$ in
the relativistic case.

-- There is no unitary time development of the system on the
quantum level.

In this paper we formulate unitary noncommutative $q$-dynamics
on the quantum level. To do this let us notice that a possible
deformation of the standard quantum mechanics lies in change of
the algebra of observables with consequences on the level of
dynamics. This is pictured on the Fig.\ 1.
\begin{figure}[t]
\begin{tabbing}
\xpt\sf
can be possibly changed:\=
\fbox{\parbox{3in}{\xpt\sf\begin{center}
\underline{\uppercase{Algebra of Observables}}\\
in the standard case generated\\
by\\
$\sf x,p,I:\;[x,p]=i\hbar I$
\end{center}}}\\
\\
\xpt\sf
unchanged:\>
\fbox{\parbox{3in}{\xpt\sf\begin{center}
\underline{\uppercase{States \& Measurement}}\\
--Notion of states\\
--Reduction of states\\
--What is measured\\
--Average values
\end{center}}}\\
\parbox{2in}{\xpt\sf probabilistic\\ interpretation\\ of QM}\>
\parbox{3in}{\centerline{$\Downarrow$}}\\
\xpt\sf cannot be changed:\>
\fbox{\parbox{3in}{\xpt\sf\begin{center}
\underline{\uppercase{Quantum Dynamics}}\\
unitary time development\\
$\Updownarrow$\\
Heisenberg equations of motion:\\
$\sf \dot{\Omega}=(i/\hbar)[H,\Omega]+\partial_t\Omega$
\end{center}}}\\
\\
\parbox{2in}{\xpt\sf a possible convenient\\ description:}\>
\fbox{\parbox{3in}{\xpt\sf\begin{center}
\underline{\uppercase{Quantum de Rham Complex}}\\
Hamilton equations \&\\
algebra of observables
\end{center}}}
\end{tabbing}
\xpt{\bf Figure 1.}
This scheme is showing possible changes in the structure of QM
\end{figure}
The main observation is  the well known statement, that
probabilistic interpretation of  quantum mechanics causes an
unitary time evolution of physical system irrespectively of the
choice of the algebra of  observables (standard or $q$-deformed).
As a consequence the Heisenberg equations of motion hold in each
case (in the Heisenberg picture). In the following we restrict
ourselves to the one degree of freedom systems.\vspace{2\baselineskip}

\noindent
{\bf ALGEBRA OF OBSERVABLES---STANDARD QM CASE}\vspace{\baselineskip}

Construction of quantum spaces by Manin$^4$ as quotient of a
free algebra by two-sided ideal can be applied also to the
Heisenberg algebra case. In fact the Heisenberg algebra can be
introduced as the quotient algebra
\begin{equation}\label{1}
{\cal H}=A(I,x.p)/J(I,x.p)
\end{equation}
where $A(I,x,p)$ is an unital associative algebra freely
generated by $I$, $x$ and $p$,  while $J(I,x,p)$ is a
two-sided ideal in $A$ defined by the Heisenberg rule
\begin{equation}\label{2}
xp=px+\i\hbar I.
\end{equation}
There is an antilinear anti-involution (star operation) in $A$
defined on generators as below
\begin{equation}\label{3}
x^*=x,\quad p^*=p.
\end{equation}
{}From the above construction it follows that this
anti-involution induces in $\cal H$ a $^*$-anti-automorphism defined again
by the eqs.\ (\ref{3}).

Now, according to the result of Aref'eva \&
Volovich$^2$, confirmed in Rembielinski$^3$ for the relativistic case,
some parameters of the considered dynamics, like inertial mass,
do not commute with the generators $x$ and $p$. This means that
these parameters should be treated themselves as generators of
the algebra. To be more concrete let us consider a conservative
system described by the Hamiltonian
\begin{equation}\label{4}
H+p^2\kappa^2+V(x,\kappa,\lambda).
\end{equation}
Here $\kappa$ and $\lambda$ are assumed to be additional
hermitean generators of the extended algebra $\cal H'$
satisfying the following re-ordering rules
\begin{eqnarray}
xp&=&px+\i\hbar\lambda^2\nonumber\\
x\lambda&=&\lambda x\nonumber\\
p\lambda&=&\lambda p\label{5}\\
x\kappa&=&\kappa x\nonumber\\
p\kappa&=&\kappa p\nonumber\\
\kappa\lambda&=&\lambda\kappa.\nonumber
\end{eqnarray}

We observe that the
generators $\kappa$ and $\lambda$ belong to the center of $\cal
H'$. Thus the irreducibility condition on the representation
level implies that $\lambda$ and $\kappa$ are multipliers of the
identity $I$.  Consequently they can be chosen as follows
\begin{eqnarray}
\lambda&=&I\nonumber\\
\kappa&=&\frac{1}{\sqrt{2\mu}}I\label{6}
\end{eqnarray}
so the extended algebra $\cal H'$ reduces to the homomorphic
Heisenberg algebra $\cal H$ defined by (\ref{1}) and (\ref{2}).
Notice that $\cal H'$ can be interpreted as a quotient
of a free unital, associative and involutive algebra
$A(I,x,p,\kappa,\lambda)$ by the two-sided
ideal $J(I,x,p,\kappa,\lambda)$ defined by eqs.\ (\ref{5}) i.e.\
\begin{equation}\label{7}
{\cal H'}=A(I,x,p,\kappa,\lambda)/J(I,x,p,\kappa,\lambda)
\end{equation}
It is remarkable, that eqs.\ (\ref{5}) are nothing
but the Bethe Ansatz for $\cal H'$.

Finally, dynamics defined by the Hamiltonian $H$ and the
Heisenberg equations lead to the Hamilton form of the equations
of motion:
\begin{eqnarray}
\dot{\lambda}&=&0\nonumber\\
\dot{\kappa}&=&0\nonumber\\
\dot{x}&=&\frac{1}{\mu}p\label{8}\\
\dot{p}&=&-V'(x).\nonumber
\end{eqnarray}
\vspace{2\baselineskip}

\noindent
{\bf ALGEBRA OF OBSERVABLES---$q$-QM CASE}\vspace{\baselineskip}

Now, the formulation of the standard quantum mechanics by means
of the algebra $\cal H'$ suggest a natural
$q$-deformation of the algebra of observables; namely the
$q$-deformed algebra ${\cal H}_q$ is a quotient algebra
\begin{equation}\label{9}
{\cal H}_q=a(I,x,p,K,{\mit\Lambda})/J(I,x,p,K,{\mit\Lambda})
\end{equation}
where the two-sided ideal $J$ is defined
now by the following Bethe Ansatz re-ordering rules
\begin{eqnarray}
xp&=&q^2px+\i\hbar q{\mit\Lambda}^2\nonumber\\
x{\mit\Lambda}&=&\xi{\mit\Lambda}x\nonumber\\
p{\mit\Lambda}&=&\xi^{-1}{\mit\Lambda}p\label{10}\\
xK&=&\tau^2Kx\nonumber\\
pK&=&\varepsilon^2Kp\nonumber\\
{\mit\Lambda}K&=&\tau\varepsilon K{\mit\Lambda}\nonumber
\end{eqnarray}
where $K$ and $\mit\Lambda$ are assumed to be invertible and
\begin{equation}\label{11}
x^*=x,\quad p^*=p,\quad K^*=K,\quad {\mit\Lambda}^*={\mit\Lambda}.
\end{equation}
A consistency of the system (\ref{10}) requires
\begin{equation}\label{12}
|q|=|\xi|=|\tau|=|\varepsilon|=1.
\end{equation}
The corresponding conservative Hamiltonian has the form
\begin{equation}\label{13}
H=p^2K^2+V(x,K,{\mit\Lambda}).
\end{equation}
Now, similary
to the standard case, $\mit\Lambda$ and $K$ are assumed
constant in time:
\begin{eqnarray}
\dot{\mit\Lambda}&=&\frac{\i}{\hbar}[H,{\mit\Lambda}]\equiv0\label{14}\\
\dot{K}&=&\frac{\i}{\hbar}[H,{\mit\Lambda}]\equiv0\label{15}
\end{eqnarray}
which implies, under the assumption of the proper classical
limit (\ref{5}),
\begin{eqnarray}
\varepsilon&=&1\label{16}\\
\tau&=&\xi^{-1}\nonumber
\end{eqnarray}
and by means of eqs.\ (\ref{16})
\begin{eqnarray}
V(x,K,{\mit\Lambda})&=&V(\xi x,\xi K,{\mit\Lambda})\label{17}\\
V(x,K,{\mit\Lambda})&=&V(\xi^2x,K,\xi{\mit\Lambda})\nonumber
\end{eqnarray}
Furthermore, taking into account (\ref{16})
\begin{equation}\label{18}
\dot{x}=\frac{\i}{\hbar}[H,x]=K^2[\frac{\i}{\hbar}{\textstyle
(1-(\frac{q}{\xi})^4)p^2x+q\xi^{-4}((\frac{q}{\xi})^2+1){\mit\Lambda}^2p}],
\end{equation}
and
\begin{eqnarray}
\lefteqn{\dot{p}=\frac{\i}{\hbar}[H,p]=
-\frac{\i}{\hbar}p[V(x,K,{\mit\Lambda})-V(q^2x,K,\xi{\mit\Lambda})]+}
\nonumber\\
&&
-\frac{q}{\textstyle(\frac{q}{\xi})^2-1}\frac{1}{x}
[{\textstyle V((\frac{q}{\xi})^2x,\xi^{-2}K,\xi{\mit\Lambda})-V(x,\xi^{-2}K,
\xi{\mit\Lambda})}]{\mit\Lambda}^2.\label{19}
\end{eqnarray}
Notice that the last term is the quantum (Gauss-Jackson)
gradient of $V(x,\xi^{-2}K,\xi{\mit\Lambda}){\mit\Lambda}^2$.

Now, a consistency of the Hamilton form of the equations of
motion (\ref{14}), (\ref{15}), (\ref{18}) and (\ref{19}) with
the algebra (\ref{10}) and with the Leibniz rule confirms
(\ref{16})--(\ref{17}) and implies
\begin{equation}
V(x,K,{\mit\Lambda})=V(({\textstyle\frac{q}{\xi}})^2x,K,{\mit\Lambda})
\label{20}
\end{equation}
Furthermore, eqs.\ (\ref{17}) and (\ref{20}) implies that in the
formula (\ref{19}) the term linear in $p$ vanish. Consequently
\begin{equation}
\dot{p}=-q{\textstyle\partial^{(q/\xi)^2}_x}
V(x,\xi^{-2}K,\xi{\mit\Lambda}){\mit\Lambda}^2
\label{21}
\end{equation}
where $\partial^{(q/\xi)^2}_x$ is the Gauss-Jackson derivative
as defined in the eq.\ (\ref{19}).

Moreover, under the assumption of the proper classical limit,
eq.\ (\ref{20}) implies that
\begin{equation}
\xi=q
\label{22}
\end{equation}
and $V$ depends only on the variable $xK^{-1}{\mit\Lambda}^{-2}$ or $V$ does
not depend on $x$, so taking into account (\ref{17}) we obtain in this
case
\begin{equation}
V=0.
\label{23}\end{equation}
Therefore we have two cases.
\vspace{\baselineskip}

\noindent
{\bf Case I}
\vspace{\baselineskip}

\begin{displaymath}
H=p^2K^2
\end{displaymath}
\begin{equation}\label{24}
\dot{x}=\left[\frac{\i}{\hbar}\left(\xi^4-q^4\right)+
q\left(\xi^2+q^2\right)p{\mit\Lambda}^2\right]K^2
\end{equation}
\begin{displaymath}
\dot{p}=0
\end{displaymath}
and
\begin{eqnarray}
xp&=&q^2px+\i\hbar q{\mit\Lambda}^2\nonumber\\
x{\mit\Lambda}&=&\xi{\mit\Lambda}x\nonumber\\
p{\mit\Lambda}&=&\xi^{-1}{\mit\Lambda}p\label{25}\\
xK&=&\xi^{-2}Kx\nonumber\\
pK&=&Kp\nonumber\\
{\mit\Lambda}K&=&\xi^{-1}K{\mit\Lambda}.\nonumber\\
\end{eqnarray}
\vspace{\baselineskip}

\noindent
{\bf Case II}
\vspace{\baselineskip}

\begin{displaymath}
H=p^2K^2+V\left((2m)^{-1/2}q^{-1}xK^{-1}{\mit\Lambda}^{-2}\right)
\end{displaymath}
\begin{equation}
\dot{x}=2({\mit\Lambda}K)^{2}p
\label{26}
\end{equation}
\begin{displaymath}
\dot{p}=-q(\partial_xV){\mit\Lambda}^2.
\end{displaymath}
and the algebra (\ref{25}) holds under the condition
(\ref{22}) $\xi=q$. The meaning of the
normalisation factor $\sqrt{2m}$, $m>0$, will be evident
later. Notice that from the eqs.\ (\ref{26}) we can identify the
inertial mass $M$ as
\begin{equation}
M={\textstyle\frac{1}{2}}q(K{\mit\Lambda})^{-2},
\label{27}
\end{equation}
so
\begin{eqnarray}
xM&=&q^2Mx\nonumber\\
pM&=&q^2Mp\label{28}\\
{\mit\Lambda}M&=&q^2M{\mit\Lambda}.\nonumber
\end{eqnarray}

Now, let us consider the dynamical models by Aref'eva \&
Volovich$^2$.
\vspace{\baselineskip}

\noindent
{\bf Free particle}
\vspace{\baselineskip}

We choose the potential $V=0$ so $H=p^2K^2$ and
consequently
\begin{equation}
\dot{x}=q^{-1}M^{-1}p\label{29}
\end{equation}
\begin{displaymath}
\dot{p}=0.
\end{displaymath}
Notice
that eqs.\ (\ref{29}) do not contain $\mit\Lambda$. The
equations (\ref{29}) and the algebra (\ref{28}) are the same as in Aref'eva
\& Volovich$^2$. However it is impossible to fulfil the unitarity
condition without of the operator $\mit\Lambda$ (rest
of the algebra is defined by eqs.\ (\ref{25}), (\ref{26}). Therefore the
lacking of the unitarity in Aref'eva \& Volovich$^2$ is caused by
the choice ${\mit\Lambda}=I$ which contradicts the
reordering rules (\ref{25}).
\vspace{\baselineskip}

\noindent
{\bf Harmonic oscillator}
\vspace{\baselineskip}

We start with the Hamiltonian:
\begin{equation}
H=p^2K^2+\frac{\omega^2}{2}(q^{-1}xK^{-1}{\mit\Lambda}^{-2})^2.
\label{30}
\end{equation}
Consequently
\begin{eqnarray}
\dot{x}&=&q^{-1}M^{-1}p\label{31}\\
\dot{p}&=&-\frac{\omega^2}{2}xM.\nonumber
\end{eqnarray}
Eqs.\ (\ref{31}) still do not contain $\mit\Lambda$. The
reason of the lacking unitarity in the Aref'eva \& Volovich$^2$ is
the same as in the free-particle case.\vspace{2\baselineskip}

\noindent
{\bf \uppercase{Reparametrisation}}\vspace{\baselineskip}

The dependence of the potential $V$ on the element
$q^{-1}(2m)^{-1/2}xK^{-1}{\mit\Lambda}^{-2}$ and the form of the
ki\-ne\-tic term in Ha\-mil\-tonian (\ref{26}) suggest the following
non-cano\-nical re\-pa\-ra\-met\-ri\-sa\-tion of the $q$-QM dynamics in the
Case II:
\begin{eqnarray}
X&:=&q^{-1}(2m)^{-1/2}xK^{-1}{\mit\Lambda}^{-2}\label{32}\\
P&:=&(2m)^{1/2}pK.\nonumber
\end{eqnarray}
By means of the eqs.\ (\ref{32}), (\ref{25}) and (\ref{22}) we
obtain the following form of the reordering rules (in terms of
$X$, $P$, $K$, and $\mit\lambda$)
\begin{eqnarray}
XP&=&PX+\i\hbar I\nonumber\\
K{\mit\Lambda}&=&q{\mit\Lambda}K\label{33}\\
\null[{\mit\Lambda},X]=[{\mit\Lambda},P]&=&[K,X]=[K,P]=0.\nonumber
\end{eqnarray}
Therefore
\begin{equation}
{\cal H}_q={\cal H}\oplus{\cal M}^2_q
\label{34}
\end{equation}
i.e.\ ${\cal H}_q$ is the direct sum of the Heisenberg algebra
generated by $X$, $P$ and of the real Manin's plane ${\cal M}^2_q$
(generated by $K$ and ${\mit\Lambda}$). Moreover the
Hamilton equations take the standard form
\begin{eqnarray}
\dot{X}&=&\frac{1}{m}P\label{35}\\
\dot{P}&=&-V'(X)\nonumber
\end{eqnarray}
with
\begin{equation}
H=p^2K^2+V(q^{-1}(2m)^{-1/2}xK^{-1}{\mit\Lambda}^{-2})=P^2\frac{1}{2m}+V(X).
\label{36}
\end{equation}
It is evident that energy spectra of both dynamics (defined by
$x$ and $p$ or by $X$ and $P$) are the same.  However both
theories are unitary nonequivalent so its physical content
(identification of observables) is rather different. In the Case
I for $\xi=q$ an analogous reparametrisation is impossible. It
is remarkable, that a similar analysis given in Brzezinski \&
al.$^5$ for a quantum particle on a $q$-circle leads to quite
analogous conclusions.\vspace{2\baselineskip}

\noindent
{\bf\uppercase{Quantum de Rham complex}}
\vspace{\baselineskip}

Now, we observe that the Hamiltonian equations of motion
(\ref{8}) in the standard quantum mechanics can be written as
\begin{eqnarray}
\d x&\equiv&\dot{x}\d t=\frac{1}{\mu}p\,\d t\label{37}\\
\d p&\equiv&\dot{p}\d t=-V'(x)\,\d t.\nonumber
\end{eqnarray}
By means of the Heisenberg reordering rule (\ref{2}) it is easy
to calculate that
\begin{eqnarray}
x\,\d x&=&\d x\,(x+\i\hbar p^{-1})\nonumber\\ p\,\d x&=&\d
x\,p\label{38}\\ x\,\d p&=&\d p\,x\nonumber\\ p\,\d p&=&\d
p\,(p-\i\hbar V''(x)/V'(x))\nonumber
\end{eqnarray}
or in a more symmetric form
\begin{eqnarray}
px\,\d x&=&\d x\,xp\nonumber\\ p\,\d x&=&\d x\,p\label{39}\\
x\,\d p&=&\d p\,x\nonumber\\ V'(x)p\,\d p&=&\d
p\,pV'(x).\nonumber
\end{eqnarray}

Assuming that $\d x$ and $\d p$ are obtained from $x$ and $p$
respectively as an effect of the application of external
differential $\d$ satisfying usual conditions (linearity,
nilpotency and the graded Leibniz rule) we can complete the
differential algebra with a two-form sector.  It is matter of
simple calculations to show that
\begin{eqnarray}
\d x\,\d p&=&-\d p\,\d x\nonumber\\
p^2(\d x)^2&=&(px-\i\hbar/2)\d x\,\d p\label{40}\\ (\d
p)^2&=&-\frac{\i\hbar}{2}\d x\,\d p\,{\rm
D}_x\left(\frac{V''(x)}{V'(x)}\right),
\nonumber
\end{eqnarray}
where ${\rm D}_x$ is the partial $\hbar$-derivative with respect
to $x$, defined via $\d f(x,p)=\d x\,{\rm D}_xf+\d p\,{\rm
D}_pf$.

As a consequence
\begin{equation}
(\d x)^3=(\d p)^3=\d x\,(\d p)^2=(\d x)^2\,\d p=0.
\label{41}
\end{equation}
Therefore we have defined a $Z_2$ graded $\cal H$-bi-module with
$\dim({\cal H})=1+2+1=4$---a quantum analogon of the deRham
complex.

Now, the above quantum deRham complex can be $q$-deformed
according to the deformation of the Heisenberg algebra $\cal H$.
The resulting first order differential calculus reads
\begin{eqnarray}
px\,\d x&=&q^{-4}\d x\,xp\nonumber\\ x\,\d p&=&q^2\d
p\,x\nonumber\\
\d x\,p&=&q^2p\,\d x\label{42}\\
\partial_xV(X)p\,\d p&=&q^{-4}\d p\,p\partial_xV(X)\nonumber\\
\d K&=&\d{\mit\Lambda}=0,\nonumber
\end{eqnarray}
where $X$ is given in (\ref{32}) while the derivative
$\partial_x$ is with respect to $x$.

It can be verified that the Hamilton equations (\ref{26}) can be
reconstructed from (\ref{42}) by means of the eqs.\ (\ref{22}),
(\ref{25}) under substitution
\begin{eqnarray}
\d x&=&\dot{x}(x,p,K,{\mit\Lambda})\d t\label{43}\\
\d p&=&\dot{p}(x,p,K,{\mit\Lambda})\d t.\nonumber
\end{eqnarray}

Therefore the quantum deRham complex contain all
information about the algebra of observables and dynamics of the
theory.

Recently Dimakis et al.$^6$ also applied some
differential geometric methods to the Heisenberg algebra but
from another point of view.
\vspace{2\baselineskip}

\noindent
{\bf \uppercase{Acknowledgements}}
\vspace{\baselineskip}

I am  grateful for interesting discussions to
Prof.~H.D.~Doebner, Prof.~W.~Tybor, Mr.~T.~Brzezinski and
Mr.~K.~Smolinski.

This work is supported under Grant KBN 2 0218 91 01.
\vspace{2\baselineskip}

\noindent
{\bf\uppercase{References}}
\vspace{\baselineskip}

\xpt
\noindent
1. L.C. Biedenharn, {\it J.  Phys.  A: Math. Gen.} 22:L873 (1989).\\
2.  I.Ya.  Aref'eva, I.V.  Volovich, Quantum group particles and
non-Archimedean geometry,\\
\indent{\it preprint} CERN--TH--6137/91 (1991).\\
3.  J.  Rembieli\'nski, {\it Phys.  Lett. B\/}287:145 (1992).\\
4. Yu.I.  Manin, ``Quantum Groups  and Non-Commutative Geometry'',
publication CRM, Montreal\\
\indent(1988).\\
5.  T. Brzezi\'nski, J.  Rembieli\'nski, K.A.  Smoli\'nski, Quantum
particle on a quantum circle,\\
\indent {\it {\L}\'od\'z University preprint} KFT U{\L}
92--10 (1992).\\
6. A.  Dimakis, F. M\"uller-Hoissen, Quantum mechanics as a
non-commutative symplectic geometry,\\
\indent {\it G\"ottingen University
preprint}, (1992).

\end{document}